\documentclass[12pt]{amsart}
\usepackage{amssymb}
\usepackage{color}
\usepackage{amsmath,epic,curves}
\usepackage[english]{babel}
\usepackage{graphicx}
\usepackage{comment}
\newtheorem{claim}{}[section]
\newtheorem{theorem}[claim]{Theorem}

\newtheorem{proposition}[claim]{Proposition}

\begin{document}
\baselineskip 6.0 truemm
\parindent 1.5 true pc

\newcommand\lan{\langle}
\newcommand\ran{\rangle}
\newcommand\tr{{\text{\rm Tr}}\,}
\newcommand\ot{\otimes}
\newcommand\ol{\overline}
\newcommand\join{\vee}
\newcommand\meet{\wedge}
\renewcommand\ker{{\text{\rm Ker}}\,}
\newcommand\image{{\text{\rm Im}}\,}
\newcommand\id{{\text{\rm id}}}
\newcommand\tp{{\text{\rm tp}}}
\newcommand\pr{\prime}
\newcommand\e{\epsilon}
\newcommand\la{\lambda}
\newcommand\inte{{\text{\rm int}}\,}
\newcommand\ttt{{\text{\rm t}}}
\newcommand\spa{{\text{\rm span}}\,}
\newcommand\conv{{\text{\rm conv}}\,}
\newcommand\rank{\ {\text{\rm rank of}}\ }
\newcommand\re{{\text{\rm Re}}\,}
\newcommand\ppt{\mathbb T}
\newcommand\rk{{\text{\rm rank}}\,}

\newcommand{\bra}[1]{\langle{#1}|}
\newcommand{\ket}[1]{|{#1}\rangle}

\title{Three-qubit entanglement witnesses
with the full spanning properties}

\author{Seung-Hyeok Kye}
\address{Department of Mathematics and Institute of Mathematics, Seoul National University, Seoul 151-742, Korea}

\thanks{partially supported by NRFK 2009-0083521.}



\begin{abstract}
We construct three-qubit entanglement witnesses with the properties
that all the partial transposes have the spanning properties. These
witnesses determine faces for separable states whose interior lies
in the interior of PPT states, and so they detect sets of PPT
entangled states with nonempty interiors. These witnesses also detect
all kinds of bi-separable pure product states. Our construction
depends on the observation that multi-partite entanglement witnesses
correspond to multi-linear maps which are positive.
\end{abstract}

\maketitle

\section{Introduction}

The separability problem is one of the most important research
topics in the whole theory of entanglement, and the criterion with
the positive partial transposes \cite{choi-ppt,peres} is quite
strong for this purpose. If we denote by $\mathbb S$ and $\mathbb T$
the convex sets of all separable and PPT states, respectively, then
the criterion tells us that the relation $\mathbb S\subset\mathbb T$
holds. These two convex sets coincide only for two-qubit and
qubit-qutrit cases \cite{choi-ppt,horo-1,woronowicz}. In order to
distinguish separable states among PPT states, we have to understand
the boundary structures of the convex set $\mathbb S$. Some parts of
boundary of $\mathbb S$ are determined by the boundary of $\mathbb
T$, but some parts of the boundary are located in the interior of
the convex set $\mathbb T$. The former case can be more easily understood than the latter,
because the boundary structures of the convex set $\mathbb T$ itself
can be described in terms of subspaces. See \cite{ha_kye_04} and
\cite{ha_kye_multi} for bi-partite and multi-partite cases,
respectively. In this context, the latter cases should be important
topics to be studied.

The boundary of $\mathbb S$ consists of exposed faces which are determined by entanglement witnesses.
An entanglement witness $W$ determines a face
of $\mathbb S$ whose interior lies in the interior of $\mathbb T$
if and only if the set of PPT entangled states detected by $W$ has a nonempty interior in $\mathbb T$
if and only if all the partial transposes of $W$ has the spanning properties. This is known for bi-partite cases
\cite{ha-kye-optimal,kye_ritsu}, and easily extended to multi-partite cases.
We will say that an entanglement witness $W$
has the {\sl full spanning} property when $W$ has this property.
Such witnesses have been constructed
for bi-partite cases. See \cite{ha-kye_JPA_2012,ha-kye_exp_Choi} for $3\otimes 3$ cases, and
\cite{chrus_exposed,chrus_Robert} for $4\otimes 4$ cases. Very recently, $2\otimes 4$ entanglement witnesses
with the full spanning properties have been constructed in \cite{ha-kye_2x4}.

The main purpose of this note is to construct three-qubit entanglement witnesses
with the full spanning properties. With these witnesses, it is easy to construct
three-qubit boundary separable states with full ranks, as it was requested in \cite{chen_full_sep}.
Such states have been constructed for the two-qutrit case \cite{ha-kye-sep-face},
and studied \cite{chen_full_sep} in detail. See also \cite{ha-kye_2x4} for such states in $2\otimes 4$ case.
It turns out that our witnesses also detect all kinds of bi-separable pure product states.

It was shown in \cite{horo-multi} that an $n$-partite entanglement witness
corresponds to a linear map from the tensor product of the first $(n-1)$ systems into the last system,
which sends separable states into positive matrices. We interpret this linear map as a multi-linear map
with $n-1$ variables whose linearization is a positive linear map with respect to
the maximal tensor product of matrix algebras in the category of function systems \cite{effros, han}.
This makes easy to construct required witnesses
and determine if they have the full spanning properties or not.

In the next section, we will discuss how entanglement witnesses
correspond to positive multi-linear maps, and
consider the full spanning properties of them which are
multi-partite version of the bi-spanning property
\cite{ha-kye-optimal}.
We also introduce the notion of completely positivity for multi-linear maps.
We construct above mentioned three-qubit
witnesses in Section 3 together with three-qubit boundary separable
states with full ranks.
We close this note to discuss other possible notions of positivity for bilinear maps.

The author is grateful to Kyung Hoon Han for useful discussion
as well as providing the references \cite{effros} and \cite{han}.

\section{Positive multi-linear maps as entanglement witnesses}

A state $\varrho$ on the Hilbert space ${\mathcal H}=\mathbb C^{d_1}\ot \mathbb C^{d_2}\ot\cdots\ot\mathbb C^{d_n}$
is called separable if it is the convex combination
$$
\varrho:=\sum_{i\in I}p_i |\xi_i\rangle\langle \xi_i|
$$
of pure product states $|\xi_i\rangle\langle \xi_i|$ with
\begin{equation}\label{pv-form}
|\xi_i\rangle=|\xi_{1i}\rangle |\xi_{2i}\rangle\cdots |\xi_{ni}\rangle\in
\mathbb C^{d_1}\ot \mathbb C^{d_2}\ot\cdots\ot\mathbb C^{d_n},\qquad i\in I.
\end{equation}
Therefore, $\varrho$ is a $d\times d$ density matrix with the dimension $d=\prod_{j=1}^n d_j$ of
the whole Hilbert space $\mathcal H$.
On the other hand, a Hermitian matrix $W$ is said to be an entanglement witness if
it is not positive (positive semi-definite) but satisfies the inequality
\begin{equation}\label{ew_def}
\lan\xi |W|\xi \ran \ge 0
\end{equation}
for every product vector $|\xi\ran=|\xi_1\ran |\xi_2\ran \cdots |\xi_n\ran$ in $\mathcal H$.
This property is also called block-positivity in bi-partite cases. See \cite{ssz}. We have
$$
\lan\xi |W|\xi\ran =\lan W, |\bar\xi\ran\lan\bar\xi|\ran
$$
with the usual bi-linear pairing $\lan A,B\ran=\tr(A^\ttt B)=\sum a_{ij}b_{ij}$ for
$A=[a_{ij}]$ and $B=[b_{ij}]$ in $M_d$, where $A^\ttt$ denotes the transpose of $A$.

For a given $d\times d$ matrix $W$ with $d=\prod_{i=1}^n d_i$, we write
$$
\begin{aligned}
W
&=\sum_{i_1,j_1}|i_1\ran\lan j_1|\ot W_{i_1,j_1}\in M_{d_1}\ot M_{d_2d_3\cdots d_n}\\
&=\sum_{i_1,j_1}\sum_{i_2,j_2}|i_1\ran\lan j_1|\otimes |i_2\ran\lan j_2|\ot W_{i_1i_2,j_1j_2}\in M_{d_1}\ot M_{d_2}\ot M_{d_3d_4\cdots d_n}\\
&=\cdots\\
&=\sum_{i_1,j_1,\dots, i_{n-1},j_{n-1}} |i_1\ran\lan j_1|\otimes \cdots\otimes |i_{n-1}\ran\lan j_{n-1}|\ot
W_{i_1\cdots i_{n-1},j_1\cdots j_{n-1}}
\end{aligned}
$$
which belongs to $M_{d_1}\ot \cdots \ot M_{d_{n-1}}\ot M_{d_n}$. We associate the multi-linear map $\phi_W$
from $M_{d_1}\times \cdots \times M_{d_{n-1}}$ into $M_{d_n}$ by
$$
\phi_W
(|i_1\ran\lan j_1|,\cdots, |i_{n-1}\ran\lan j_{n-1}|)= W_{i_1\cdots i_{n-1},j_1\cdots j_{n-1}}\in M_{d_n}.
$$
Conversely, for a multi-linear map $\phi$ from $M_{d_1}\times \cdots \times M_{d_{n-1}}$ into $M_{d_n}$, we associate a matrix
$W_\phi\in M_d$ by
$$
W_\phi
=\sum_{i_1,j_1,\dots, i_{n-1},j_{n-1}}
 |i_1\ran\lan j_1|\otimes \cdots\otimes |i_{n-1}\ran\lan j_{n-1}|\ot
 \phi(|i_1\ran\lan j_1|,\cdots, |i_{n-1}\ran\lan j_{n-1}|).
 $$
The correspondences $\phi\mapsto W_\phi$ and $W\mapsto \phi_W$ are nothing but the Choi-Jamio\l kowski isomorphisms \cite{choi75-10,jami}
for bi-partite case of $n=2$.

For a given multi-linear map $\phi$, we have
$$
\begin{aligned}
\lan\xi_1|\cdots \lan\xi_n
& |W_\phi|\xi_1\ran\cdots |\xi_n\ran\\
&=\sum \lan\xi_1|i_1\ran\lan j_1|\xi_1\ran \cdots
 \lan\xi_{n-1}|i_{n-1}\ran\lan j_{n-1}|\xi_{n-1}\ran \cdot\\
&\phantom{XXXXXXXXXXXX} \lan\xi_n| \phi(|i_1\ran\lan j_1|,\cdots, |i_{n-1}\ran\lan j_{n-1}|) |\xi_n\ran\\
&=\lan \xi_n \,|\, \phi( |\bar\xi_1\ran\lan\bar\xi_1 |, \cdots, |\bar\xi_{n-1}\ran\lan\bar\xi_{n-1}|) \,|\,\xi_n\ran,
\end{aligned}
$$
and so, we see that $W_\phi$ satisfies the condition (\ref{ew_def}) if and only if
$\phi$ is {\sl positive}, that is, $\phi(x_1,\dots,x_{n-1})$ is positive whenever all of
$x_1,\dots, x_{n-1}$ are positive. We denote by $\mathbb P$ the convex cone of all positive multi-linear maps.
We note that a bi-linear map $\phi$ from the product $V_1\times V_2$ of functions systems into a function system $V_3$
is positive in this sense
if and only if its linearization from $V_1\otimes_{\max} V_2$ into $V_3$ is a positive linear map,
where $V_1\otimes_{\max} V_2$ is the the maximal tensor product of function systems. See \cite{effros,han}.

We define the bilinear pairing $\lan \varrho, \phi\ran$ for $\varrho\in M_d$ and a $(n-1)$-linear map $\phi$ by
$$
\lan\varrho,\phi\ran = \lan \varrho,W_\phi\ran = \tr (W_\phi \varrho^\ttt).
$$
Then the above discussion tells us that
$\phi\in\mathbb P$ if and only if $\lan\varrho,\phi\ran\ge 0$
for every $\varrho\in\mathbb S$. By the separation theorem for a point outside of a closed convex set, we have the dual
statement as follows:

\begin{theorem}
An $n$-partite state $\varrho$ is separable if and only if $\lan\varrho,\phi\ran\ge 0$ for
every positive $(n-1)$-linear map $\phi$.
\end{theorem}

For a given positive $(n-1)$-linear map $\phi$, the set $\phi^\prime$ defined by
$$
\phi^\prime=\{\varrho\in\mathbb S: \lan\varrho,\phi\ran=0\}
$$
is a face of the convex set $\mathbb S$. Extreme points of the convex set $\phi^\prime$ are determined by
the set $P[\phi]$ of product vectors $|\xi\ran = |\xi_1\ran \cdots |\xi_n\ran$ satisfying the relation
\begin{equation}\label{def_p}
\lan \xi |W_\phi |\xi\ran = \lan W_\phi, |\bar\xi\ran\lan\bar\xi|\ran
= \lan \xi_n \,|\, \phi( |\bar\xi_1\ran\lan\bar\xi_1 |, \cdots, |\bar\xi_{n-1}\ran\lan\bar\xi_{n-1}|) \,|\,\xi_n\ran =0.
\end{equation}
We also consider the face $\phi^{\prime\prime}$ of $\mathbb P$ consisting of
$\psi\in\mathbb P$ such that $\lan\varrho,\psi\ran=0$ for each $\varrho\in\phi^\prime$.
We note that $\phi^{\prime\prime}$ is the smallest exposed face of $\mathbb P$ containing $\phi$.
Then we have the following:
$$
\psi\in\phi^{\prime\prime}
\ \Longleftrightarrow\
\lan\xi|W_\psi|\xi\ran=0\quad {\text{\rm for each}}\ \xi\in P[\phi].
$$
We call a multi-linear map $\phi$ {\sl completely positive} if $W_\phi$ is a positive matrix.
This is the case if and only if the corresponding linear map from $M_{d_1}\ot\cdots\ot M_{d_{n-1}}$ into $M_{d_n}$
induced by $\phi$ is completely positive by \cite{choi75-10}.
From the above discussion, we have the following:

\begin{proposition}\label{spanning--}
For a given positive multi-linear map $\phi$, the following are equivalent:
\begin{enumerate}
\item[(i)]
The set $P[\phi]$ of product vectors spans the whole space $\mathbb C^d$.
\item[(ii)]
The smallest exposed face of $\mathbb P$ containing $\phi$ has no completely positive map.
\end{enumerate}
\end{proposition}

We note that matrix algebras are nuclear as operator systems, and so
the maximal tensor product and the minimal tensor product of matrix algebras coincide
in the category of operator systems. Therefore, the above notion of complete positivity
may be naturally applied for multi-linear maps in general operator systems whose linearization 
is completely positive
with respect to the maximal tensor products.
See \cite{paulsen}, where the authors call those maps \lq\lq jointly completely positive\rq\rq.
See also \cite{pis_2002} for discussion on related terminologies for operator spaces.
It should be noted that the maximal tensor product of matrix algebras are different from
the minimal tensor product in the category of function systems.


We say that $\phi$ (or the corresponding Hermitian matrix $W_\phi$) has the {\sl spanning property}
when it satisfies the above conditions in Proposition \ref{spanning--}.
The notion of spanning property for bi-partite case has been introduced
in \cite{lew00}  in the context of optimality for entanglement witnesses.
Our discussion above is nothing but the multi-partite analogue.
It should be noted that even an indecomposable
extreme positive map may not have the spanning property, as the
Choi map shows. See \cite{ha-kye-optimal}.

Now, we turn our attention
to the notion of partial transposes of multi-partite states.
For a given subset $S$ of $\{1,2,\dots,n\}$, we define the linear map  $T(S)$
from $\bigotimes_{j=1}^n M_{d_j}$ into itself by
$$
(A_1\ot A_2\ot\cdots\ot A_n)^{T(S)}:=B_1\ot B_2\ot\cdots\ot B_n,
\quad \text{\rm with}\ B_j=\begin{cases} A_j^\ttt, &j\in S,\\ A_j,
&j\notin S.\end{cases}
$$
A state $\varrho$ is said to be of PPT if $\varrho^{T(S)}$ is positive for every subset
$S$. The convex set $\mathbb T$ of all PPT states is the intersection of convex sets
$$
\mathbb T^S=\{\varrho: \varrho^{T(S)} \ {\text{\rm is positive}}\}
$$
through subsets $S$ of $\{1,2,\dots,n\}$.
Faces of $\mathbb T$ can be described in terms of subspaces
\cite{ha_kye_multi}. Especially, a PPT state $\varrho$ is in the interior of $\mathbb T$ if and only if
$\varrho^{T(S)}$ has full rank for each subset $S$.

For a product vector $\ket \xi=\ket{\xi_1}\ot\cdots\ot\ket{\xi_n}$,
we also define the partial conjugate $\ket \xi^{\Gamma(S)}$ by
$$
(\ket{\xi_1}\ot\cdots\ot\ket{\xi_n})^{\Gamma(S)}
:=\ket{\eta_1}\ot\cdots\ot\ket{\eta_n}, \quad \text{\rm with}\
\ket{\eta_j}=\begin{cases} \ket{\bar{\xi}_j}, &j\in S,\\
\ket{\xi_j}, &j\notin S,\end{cases}
$$
where $|\bar \xi\rangle$ denotes the conjugate of $|\xi\rangle$.
We say that a positive multi-linear map $\phi$ has the {\sl full spanning} property
if $W_\phi^{T(S)}$ has the spanning property for every subset $S$. This is equivalent to the condition
\begin{equation}\label{spanning}
\spa \{|\xi\ran ^{\Gamma(S)}: \xi\in P[\phi]\} = \mathcal H,\quad
{\text{\rm for each subset}}\ S\subset \{1,2,\dots, n\}.
\end{equation}
We also say that $\phi$ is {\sl decomposable} if it is in the convex hull of
$$
\{\phi\in\mathbb P: W_\phi^{T(S)} \ {\text{\rm is positive}} \}
$$
through subsets $S$ of $\{1,2,\dots,n\}$. It is clear that $\phi$ is decomposable if and only if
$\lan\varrho,\phi\ran\ge 0$ for every $\varrho\in\mathbb T$.

We are ready to state the main theorem in this section. The proof
will be omitted, because it is simple and almost identical to the
bi-partite cases, for which we refer to Chapter 8 of the survey note
\cite{kye_ritsu}. Especially, a map with the full spanning property
must be indecomposable. See Figure 1 for the surrounding geometry of
separable and PPT states.

\begin{theorem}\label{main}
Let $\phi$ be a positive multi-linear map. Then the
following are equivalent:
\begin{enumerate}
\item[(i)]
$\phi$ has the full spanning property.
\item[(ii)]
The interior of the face $\phi^\prime$ lies in the interior of $\mathbb T$.
\item[(iii)]
The smallest exposed face $\phi^{\prime\prime}$ containing $\phi$ has no decomposable map.
\item[(iv)]
The set $\{\varrho \in\mathbb T: \lan\varrho,\phi\ran < 0\}$ of PPT entangled states detected by $\phi$
has a nonempty interior in $\mathbb T$.
\end{enumerate}
\end{theorem}

\bigskip
\begin{figure}[h]
\setlength{\unitlength}{1 mm}
\begin{center}
\begin{picture}(50,30)
    \qbezier(39.142,0.858)(45.000,6.716)(45.000,15.000)
  \qbezier(45.000,15.000)(45.000,23.284)(39.142,29.142)
\drawline(10,15)(45,15) \drawline(35,5)(35,25)
\put(10,15){\circle*{1}} \put(35,15){\circle*{1}}
\put(45,15){\circle*{1}} \put(8,16){$\mathbb I$}
\put(32,17){$\varrho$}
\put(32.5,7){$\mathbb S$} \put(41,7){$\mathbb T$}
\end{picture}
\end{center}
\caption{The vertical line segment represents the face $\phi^\prime$ of
$\mathbb S$ determined by the positive map $\phi$ with the full spanning property.
The points $\mathbb I$ and $\varrho$ represent
an interior point of $\mathbb S$ and
a boundary separable state with full ranks, respectively. The line segment
from $\mathbb I$ to $\varrho$ can be extended within the convex set $\mathbb T$, because
$\varrho$ is an interior point of $\mathbb T$. The curve
represents the boundary of the convex set $\mathbb T$.}
\end{figure}
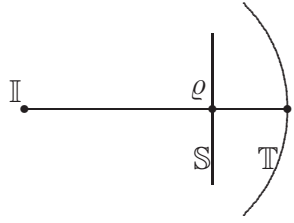
\bigskip

\section{Construction for the three-qubit case}

We begin with the following bilinear map $\phi$ from $M_2\times M_2$ into $M_2$
which sends $([x_{ij}], [y_{ij}])\in M_2\times M_2$ to
\begin{equation}\label{def_phi}
\left(\begin{matrix}
sx_{22}y_{11}&
x_{12}y_{12}-x_{12}y_{21}+x_{21}y_{12}+x_{21}y_{21}\\
x_{12}y_{12}+x_{12}y_{21}-x_{21}y_{12}+x_{21}y_{21}
&t x_{11}y_{22}
\end{matrix}\right),
\end{equation}
where $s,t$ are positive numbers with the relation $st=8$.
The corresponding entanglement witness is given by
$$
W_\phi=\left(\begin{matrix}
\cdot&\cdot&\cdot&\cdot&\cdot&\cdot&\cdot &1\\
\cdot&\cdot&\cdot&\cdot&\cdot&\cdot&1&\cdot\\
\cdot&\cdot&\cdot&\cdot&\cdot&-1&\cdot&\cdot\\
\cdot&\cdot&\cdot&t&1&\cdot&\cdot&\cdot\\
\cdot&\cdot&\cdot&1&s&\cdot&\cdot&\cdot\\
\cdot&\cdot&-1&\cdot&\cdot&\cdot&\cdot&\cdot\\
\cdot&1&\cdot&\cdot&\cdot&\cdot&\cdot&\cdot\\
1&\cdot&\cdot&\cdot&\cdot&\cdot&\cdot&\cdot
\end{matrix}\right),
$$
where $\cdot$ denotes zero.

In order to show that $\phi$ is a positive bilinear map, we consider
the rank one projection
$P_\alpha=\left(\begin{matrix}1&\bar\alpha\\
\alpha&|\alpha|^2\end{matrix}\right)$. We see that $\phi$ sends
$(P_\alpha,P_\beta)$ to
$$
\left(\begin{matrix}
s|\alpha|^2&
\alpha\beta-\bar\alpha\beta+\alpha\bar\beta+\bar\alpha\bar\beta\\
\alpha\beta+\bar\alpha\beta-\alpha\bar\beta+\bar\alpha\bar\beta
&t|\beta|^2
\end{matrix}\right)
$$
whose determinant is given by the nonnegative number
\begin{equation}\label{det}
D=|\alpha\beta-\bar\alpha\bar\beta|^2
+|\alpha\bar\beta+\bar\alpha\beta|^2.
\end{equation}

We proceed to determine the product vectors in $P[\phi]$. First of all,
we look at the image under $\phi$ when one of variables is
$P_0=|0\ran\lan 0|$ or
$P_\infty=|1\ran\lan 1|$ as follows:
$$
\begin{aligned}
\phi(|0\ran\lan 0|,y)=\left(\begin{matrix}0&0\\0&t y_{22}\end{matrix}\right),
\quad
&\phi(|1\ran\lan 1|,y)=\left(\begin{matrix}s y_{11}&0\\0&0\end{matrix}\right)\\
\phi(x,|0\ran\lan 0|)=\left(\begin{matrix}s x_{22}&0\\0&0\end{matrix}\right),
\quad
&\phi(x,|1\ran\lan 1|)=\left(\begin{matrix}0&0\\0& t x_{11}\end{matrix}\right).
\end{aligned}
$$
We see that
$$
\begin{aligned}
\lan\zeta |\phi(|0\ran\lan 0|,y)|\zeta\ran=0\
&\Longleftrightarrow\ |\zeta\ran=|0\ran\
{\text{\rm or}}\ y=|0\ran\lan 0|\\
\lan\zeta |\phi(|1\ran\lan 1|,y)|\zeta\ran=0\
&\Longleftrightarrow\ |\zeta\ran=|1\ran\
{\text{\rm or}}\ y=|1\ran\lan 1|\\
\lan\zeta |\phi(x,|0\ran\lan 0|)|\zeta\ran=0\
&\Longleftrightarrow\ |\zeta\ran=|1\ran\
{\text{\rm or}}\ x=|0\ran\lan 0|\\
\lan\zeta |\phi(x,|1\ran\lan 1|)|\zeta\ran=0\
&\Longleftrightarrow\ |\zeta\ran=|0\ran\
{\text{\rm or}}\ x=|1\ran\lan 1|.
\end{aligned}
$$
From these, we found the following product vectors in $P[\phi]$:
\begin{equation}\label{pv_1}
|\xi\ran 0\ran 1\ran,\quad
|\xi\ran 1\ran 0\ran,\quad
|0\ran \eta\ran 0\ran,\quad
|1\ran \eta\ran 1\ran,\quad
|0\ran 0\ran \zeta\ran,\quad
|1\ran 1\ran \zeta\ran,
\end{equation}
with arbitrary $|\xi\ran,|\eta\ran,|\zeta\ran\in\mathbb C^2$.
We note that product vectors in (\ref{pv_1}) span the $6$-dimensional subspace
of $\mathbb C^2\otimes\mathbb C^2\otimes \mathbb C^2$ whose orthogonal
complement is spanned by $|0\ran 1\ran 1\ran$ and $|1\ran 0\ran 0\ran$.
We also note that these product vectors are parameterized by six Riemann spheres.
If we put these six Riemann spheres in a suitable cyclic way then two adjacent
Riemann spheres have exactly one common point. See Figure 2.

\bigskip
\begin{figure}[h]
\setlength{\unitlength}{1.2 cm}
\begin{center}
\begin{picture}(5,4.3)
\thicklines
\drawline(1,0.3)(3,0.3)
\drawline(3,0.3)(4,2.03)
\drawline(4,2.03)(3,3.76)
\drawline(3,3.76)(1,3.76)
\drawline(1,3.76)(0,2.03)
\drawline(0,2.03)(1,0.3)
\put(1.5,-0.1){$|\xi\ran 1\ran 0\ran$}
\put(-0.6,0.9){$|0\ran\eta\ran 0\ran$}
\put(3.6,0.9){$|1\ran 1\ran \zeta\ran$}
\put(3.6,2.9){$|1\ran \eta\ran 1\ran$}
\put(-0.6,2.9){$|0\ran 0\ran \zeta\ran$}
\put(1.5,3.9){$|\xi\ran 0\ran 1\ran$}

\put(1,0.3){\circle*{0.1}}
\put(3,0.3){\circle*{0.1}}
\put(4,2.03){\circle*{0.1}}
\put(3,3.76){\circle*{0.1}}
\put(1,3.76){\circle*{0.1}}
\put(0,2.03){\circle*{0.1}}

\put(3.1,0.1){$|110\ran$}
\put(4.1,1.93){$|111\ran$}
\put(3.1,3.7){$|101\ran$}
\put(0.1,3.7){$|001\ran$}
\put(-0.9,1.93){$|000\ran$}
\put(0.1,0.1){$|010\ran$}

\end{picture}
\end{center}
\caption{Each line segment represents the Riemann sphere which parameterizes product vectors
with a single variable.}
\end{figure}
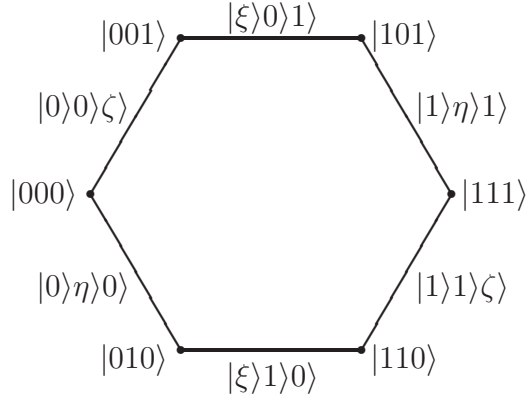
\bigskip

To find further product vectors
in $P[\phi]$, we solve the equation $D=0$ with (\ref{det}) to get the
nonzero solutions $(\alpha,\beta)$:
$$
(a_1\omega,b_1\omega^7),\qquad
(a_2\omega^3,b_2\omega^5),\qquad
(a_3\omega^5,b_3\omega^3),\qquad
(a_4\omega^7,b_4\omega),
$$
where $a_i,b_i$ are arbitrary positive numbers and
$\omega=e^{\frac{\pi i}4}$ is the
$8$-th root of unity. The corresponding $2\times 2$ matrices
$\phi(P_\alpha, P_\beta)$ are given by
$$
\begin{aligned}
\left(\begin{matrix}s a_1^2&ra_1b_1\omega\\
r a_1b_1\omega^7&tb_1^2\end{matrix}\right),
\qquad
&\left(\begin{matrix}s a_2^2&ra_2b_2\omega^7\\
r a_2b_2\omega&tb_2\end{matrix}\right),
\\
\left(\begin{matrix}s a_3^2&ra_3b_3\omega\\
r a_3b_3\omega^7&tb_3\end{matrix}\right),
\qquad
&\left(\begin{matrix}s a_4^2&ra_4b_4\omega^7\\
r a_4b_4\omega&tb_4^2\end{matrix}\right),
\end{aligned}
$$
where $r=2\sqrt 2$. The corresponding kernel vectors are also given by
$$
(rb_1,sa_1\omega^3)^\ttt,\qquad
(rb_2,sa_2\omega^5)^\ttt,\qquad
(rb_3,sa_3\omega^3)^\ttt,\qquad
(rb_4,sa_4\omega^5)^\ttt.
$$
By (\ref{def_p}), we have the following product vectors
in $P[\phi]$:
\begin{equation}\label{pv_2}
\begin{aligned}
\zeta_1(a_1,b_1)
&=(1,a_1\omega^7)^\ttt\otimes (1,b_1\omega)^\ttt\otimes (b_1,ua_1\omega^3)^\ttt
\\
\zeta_2(a_2,b_2)
&=(1,a_2\omega^5)^\ttt\otimes (1,b_2\omega^3)^\ttt\otimes (b_2,ua_2\omega^5)^\ttt
\\
\zeta_3(a_3,b_3)
&=(1,a_3\omega^3)^\ttt\otimes (1,b_3\omega^5)^\ttt\otimes (b_3,ua_3\omega^3)^\ttt
\\
\zeta_4(a_4,b_4)
&=(1,a_4\omega)^\ttt\otimes (1,b_4\omega^7)^\ttt\otimes (b_4,ua_4\omega^5)^\ttt
\end{aligned}
\end{equation}
where $u=\dfrac sr=\dfrac s{2\sqrt 2}$. Therefore, the set $P[\phi]$ consists of
product vectors in (\ref{pv_1}) and (\ref{pv_2}). It is trivial to check
the condition (\ref{spanning}) for the full spanning property of $\phi$.

\begin{theorem}
For any positive numbers $s,t$ with $st=8$, the map $\phi:M_2\times
M_2\to M_2$ defined by {\rm (\ref{def_phi})} is a
positive bilinear map with the full spanning property.
\end{theorem}

Typical PPT entangled states detected by $\phi$ are so called $X$-shaped states
\cite{acin,vin,wein}. For example, we consider the following state
$$
\varrho=\left(\begin{matrix}
1 &\cdot &\cdot &\cdot &\cdot &\cdot &\cdot &-1\\
\cdot  &1 &\cdot &\cdot &\cdot &\cdot &-1 &\cdot\\
\cdot  &\cdot &1 &\cdot &\cdot &1 &\cdot &\cdot\\
\cdot  &\cdot &\cdot &s/2\sqrt 2 &-1 &\cdot &\cdot &\cdot\\
\cdot  &\cdot &\cdot &-1 &t/2\sqrt 2 &\cdot &\cdot &\cdot\\
\cdot  &\cdot &1 &\cdot &\cdot &1 &\cdot &\cdot\\
\cdot  &-1 &\cdot &\cdot &\cdot &\cdot & 1 &\cdot\\
-1  &\cdot &\cdot &\cdot &\cdot &\cdot &\cdot &1
\end{matrix}\right)
$$
It is clear that $\varrho$ is of PPT whenever $st=8$, but we have
$$
\lan \varrho,\phi\ran = \frac{st}{2\sqrt 2}+\frac{st}{2\sqrt 2}-8=\frac 8{\sqrt 2}-8<0,
$$
as it is required.

In order to look for bi-separable states detected by $\phi$, we consider
$$
\begin{aligned}
\xi_1(\alpha)
&=(1,\bar\alpha)^\ttt_A\otimes (0,1,\alpha,0)^\ttt_{BC}
=(0,1,\alpha,0,0,\bar\alpha,|\alpha|^2,0)^\ttt\\
\xi_2(\alpha)
&=(1,\bar\alpha)^\ttt_B\otimes (1,0,0,\alpha)^\ttt_{AC}
=(1,0,\bar\alpha,0,0,\alpha,0,|\alpha|^2)^\ttt\\
\xi_3(\alpha)
&=(1,\bar\alpha)^\ttt_C\otimes (1,0,0,\alpha)^\ttt_{AB}
=(1,\bar\alpha,0,0,0,0,\alpha,|\alpha|^2)^\ttt
\end{aligned}
$$
Then we see that
$$
\lan W_\phi, |\xi_i(\alpha)\ran\lan \xi_i(\alpha)|\ran
= -2|\alpha|^2\pm (\alpha^2+\bar\alpha^2) <0
$$
for each $i=1,2,3$, whenever $\alpha^2\in\mathbb C\setminus\mathbb R$.

We proceed to search boundary separable states with full ranks.
For simplicity, we fix $s=t=2\sqrt 2$ in the definition of $\phi$. We
note that the following ten product vectors
$$
|000\ran,\quad
|001\ran,\quad
|010\ran,\quad
|101\ran,\quad
|110\ran,\quad
|111\ran,\quad
\zeta_i(1,1),\quad i=1,2,3,4
$$
in $P[\phi]$ span the whole space. All kinds of partial conjugates
of them also span the whole space, and so any nontrivial convex combination of the ten corresponding
pure product states must be boundary separable states with full ranks.
For more concrete examples, we denote by $\varrho_0$ the separable states
given by first six product vectors:
$$
\varrho_0=
\dfrac 16\left(\begin{matrix}
1&\cdot&\cdot&\cdot&\cdot&\cdot&\cdot&\cdot\\
\cdot&1&\cdot&\cdot&\cdot&\cdot&\cdot&\cdot\\
\cdot&\cdot&1&\cdot&\cdot&\cdot&\cdot&\cdot\\
\cdot&\cdot&\cdot&\cdot&\cdot&\cdot&\cdot&\cdot\\
\cdot&\cdot&\cdot&\cdot&\cdot&\cdot&\cdot&\cdot\\
\cdot&\cdot&\cdot&\cdot&\cdot&1&\cdot&\cdot\\
\cdot&\cdot&\cdot&\cdot&\cdot&\cdot&1&\cdot\\
\cdot&\cdot&\cdot&\cdot&\cdot&\cdot&\cdot&1
\end{matrix}\right)
$$
We also denote by $\varrho_1$ the separable state given by the last four product vectors:
$$
\varrho_1=
\dfrac 1{8\sqrt 2}
\left(\begin{matrix}
\sqrt 2 &-1 &\cdot&\cdot&\cdot&\cdot &\sqrt 2 &-1\\
-1 &\sqrt 2 &\cdot&\cdot&\cdot&\cdot &-1 &\sqrt 2\\
\cdot&\cdot  &\sqrt 2 &-1 &\cdot &1 &\cdot&\cdot\\
\cdot&\cdot  &-1 &\sqrt 2 &-1 &\cdot &\cdot&\cdot\\
\cdot&\cdot  &\cdot &-1 &\sqrt 2 &-1 &\cdot&\cdot\\
\cdot&\cdot  &1 &\cdot &-1 &\sqrt 2 &\cdot&\cdot\\
\sqrt 2 &-1 &\cdot&\cdot&\cdot&\cdot &\sqrt 2 &-1\\
-1 &\sqrt 2 &\cdot&\cdot&\cdot&\cdot &-1 &\sqrt 2
\end{matrix}\right)
$$
Then we see that
$$
\varrho_\lambda =(1-\lambda)\varrho_0+\lambda\varrho_1
$$
is a boundary separable state with full ranks whenever $0<\lambda<1$.

\section{Discussion}

In this note, we have considered two notions of positivity for multi-linear maps,
whose linearizations satisfy natural positivity with respect to the maximal tensor products
in the categories of functions systems and operator systems, respectively.

We may consider various other notions for positivity of multi-linear maps. We list up some candidates
for bilinear maps $\phi:M_A\times M_B\to M_C$:
\begin{itemize}
\item
$y\mapsto \phi(x,y):M_B\to M_C$ is completely positive for each positive $x\in M_A$.
\item
$x\mapsto \phi(x,y):M_A\to M_C$ is completely positive for each positive $y\in M_B$.
\item
The corresponding linear map $M_A\ot M_B\to M_C$ is positive.
\end{itemize}
The above properties have obvious dual objects in the contexts of bi-separable states \cite{acin}.
For example, $\phi$ satisfies the first condition if and only if the corresponding linear map $M_A\to M_B\ot M_C$
is positive if and only if $\lan\varrho,\phi\ran\ge 0$ for each $A$-$BC$ bi-separable states $\varrho$.
Therefore, a tri-partite entanglement witness detects an entangled state which is not bi-separable
if and only if the corresponding bilinear map satisfies the above three properties simultaneously.
It would be interesting to formulate the above properties in the framework of general function systems and/or
operator systems.

\end{document}